\documentclass[11pt]{article}
\usepackage[a4paper, total={6.7in, 9.5in}]{geometry}
\usepackage{graphicx}
\usepackage{subcaption}
\usepackage{booktabs} 
\usepackage{hyperref}
\usepackage{bbm}
\usepackage{tikz}
\usepackage{xcolor}
\usepackage{pifont}
\usepackage{float}
\usepackage{amsmath}
\usepackage{amssymb}
\usepackage{mathtools}
\usepackage{amsthm}
\usepackage{mathrsfs, dsfont}
\usepackage{tikz-cd}
\usepackage{tikz}
\usetikzlibrary{bayesnet,positioning}
\usepackage{pgfplots}
\pgfplotsset{compat=1.17}
\usepackage{bm}
\usepackage{natbib}
\usepackage{algorithm}
\usepackage{algorithmic}
\usepackage{caption}
\theoremstyle{plain}

\theoremstyle{definition}

\theoremstyle{remark}

\newcommand{\altprob}[1]{\mathbb{P}}

\usetikzlibrary{decorations.pathreplacing}

\usepackage[utf8]{inputenc}
\usepackage{amsmath}
\usepackage{graphicx}
\usepackage{amsthm}
\usepackage{amsfonts}
\usepackage{enumitem}
\usepackage{multirow}
\usepackage[colorinlistoftodos]{todonotes}

\usepackage{adjustbox}
\usepackage{algorithm}
\usepackage{amsmath, amsthm, amsfonts, amssymb}
\usepackage{mathrsfs}

\usepackage{cancel}
\usepackage[scale=2]{ccicons}
\usepackage{color}
\usepackage{csquotes}

\usepackage{enumitem}
\usepackage{etoolbox}
\usepackage{expl3}

\usepackage{float}
\usepackage{booktabs}

\usepackage{graphics, graphicx}

\usepackage{hyperref}
\hypersetup{colorlinks=true,allcolors=[rgb]{0,0,0}}

\usepackage[utf8]{inputenc}

\usepackage{mathtools}
\usepackage{mathtools}
\usepackage[most]{tcolorbox}
\usepackage[table]{xcolor}

\usepackage{natbib}

\usepackage{pgfpages}
\usepackage{pifont}
\usepackage[section]{placeins}

\usepackage{ragged2e}

\usepackage{sansmathaccent}
\usepackage{scrextend}
\usepackage{soul}
\usepackage{subcaption}

\usepackage{tikz}
\usetikzlibrary{positioning, backgrounds, matrix, 3d}
\usetikzlibrary{decorations.pathreplacing}
\usetikzlibrary{bayesnet}
\usetikzlibrary{arrows}
\usetikzlibrary{matrix}
\usepackage{tikz-3dplot}
\usepackage{tikz-cd}
\usepackage{tikz-network}
\usepackage{tcolorbox}

\usepackage{xcolor}
\usepackage{xparse}

\newcommand{\btheta}{ \boldsymbol{\theta} }

\newcommand{\bR}{ {\bf R} }
\newcommand{\bV}{ {\bf V} }

\newcommand{\bY}{ {\bf Y} }

\newcommand{\R}{\mathbb{R}}
\newcommand{\N}{\mathbb{N}}

\newcommand{\blind}{1}

\date{}

\title{\bf Laplace and skew-Laplace approximations 
\\
for Dirichlet process mixture posterior density}

\if0\blind
\author{}
\else
\author{
Beatrice Franzolini,\\
Department of Economics, Management and Statistics,\\ University of Milano-Bicocca
\vspace*{0.2cm}\\
Francesco Pozza\\
Bocconi Institute for Data Science and Analytics,\\ Bocconi University
}
\fi

\begin{document}
\maketitle

\bigskip
\begin{abstract}
Posterior inference for Dirichlet process mixture models is analytically intractable and typically relies on Markov chain Monte Carlo methods, which can become computationally prohibitive at moderate to large sample sizes.  
In this work, we investigate the performance of Laplace and skew-Laplace posterior approximations for density estimation in this setting. Through an extensive numerical study covering four simulation scenarios with sample sizes ranging from $n = 20$ to $n = 2{,}000$ and four standard real datasets, we compare the standard Laplace approximation, its skew-corrected extension, and a slice sampling benchmark, assessing accuracy through total variation distance and computational efficiency through runtime.  
Our results show that the Gaussian Laplace approximation is more effective in this setting than might be anticipated, and that the skew-Laplace approximation consistently improves posterior recovery while remaining substantially faster than state-of-the-art Markov chain Monte Carlo samplers across all settings considered.  
In particular, the use of skew-Laplace in place of the standard Laplace approximation is especially beneficial in more complex density structures, where we observe error reductions typically on the order of 30\%.
\end{abstract}

\vfill 
\vfill

\maketitle

\section{Introduction}\label{sec:intro}

Dirichlet process mixture (DPM) models \citep{lo1984class} are among the most widely used tools in Bayesian nonparametrics for density estimation, clustering, and flexible model-based inference. 
Their appeal stems from the ability to accommodate complex distributional shapes without requiring a finite-dimensional parametric family or a fixed number of mixture components to be specified in advance. 
This flexibility has made DPM models a standard choice in a broad range of applications, from density estimation and unsupervised learning to more structured hierarchical modeling. 
At the same time, such modeling richness comes at a substantial computational price, since posterior inference is analytically intractable and must typically be carried out through expensive simulation-based methods.

Posterior computation for DPM models has traditionally, and still largely, relied on Markov chain Monte Carlo (MCMC) methods, including marginal algorithms \citep{escobar1995bayesian,neal2000markov}, blocked Gibbs samplers \citep{ishwaran2001gibbs}, and slice samplers \citep{walker2007sampling,kalli2011slice}. 
These procedures are flexible and can deliver accurate inference, but they often come with a substantial computational cost, especially as the sample size increases, and may become prohibitively time-consuming even for moderately sized datasets. 
In addition to simulation-based methods, posterior inference for Dirichlet process mixture models has also been approached through variational Bayes. 
Beginning with the stick-breaking variational formulation of \cite{blei2006variational}, variational methods have provided a computationally attractive alternative to MCMC, offering considerable gains in speed and scalability. 
Their computational appeal, however, is achieved through restrictive approximating families which, especially in the mean-field setting, may compromise the faithful representation of posterior dependence and asymmetry \citep{blei2017variational}.

Against this background, our study is motivated by the need for a computational strategy that retains the speed of deterministic approximation while improving its ability to capture important features of the posterior distribution, most notably dependence and asymmetry. 
In this respect, skew distributions arising from the perturbation of a symmetric baseline density \citep{azzalini1985class,azzalini1996multivariate,azzalini1999statistical, azzalini2003distributions, ma2004flexible, genton2004skew, genton2005generalized, Azzalini_2013} provide a natural and tractable family for enriching Gaussian approximations. 
More broadly, recent work in approximate Bayesian inference has shown that explicit skew corrections can substantially improve Gaussian posterior approximations \citep{durante2024skewed,katsevich2023laplace,pozza2026skew}. 
At the same time, the success of Laplace-based deterministic approximation methods in other areas of Bayesian computation, most notably within the INLA framework for latent Gaussian models \citep{rue2009approximate,rue2017bayesian}, has shown that accurate alternatives to MCMC can often be constructed in highly structured settings. 
Finite mixture models have also been approached through INLA-based hybrid or conditional schemes \citep[see Chapter 13 of][]{gomez2020bayesian}. 

By contrast, Gaussian approximations have played only a limited role in the nonparametric Bayesian literature on mixture models. 
One possible reason is that the posterior structure of DPM models may appear unfavorable to Gaussian approximation, owing to asymmetry and effective multimodality induced by clustering uncertainty and label-switching phenomena \citep{hastie2015sampling}. 

In this paper, we investigate Laplace-based posterior approximation \citep[see e.g.,][Ch.~13]{gelman2013bayesian} in the DPM setting and develop an approximate sampling scheme for posterior density estimation based on the method proposed by \citet{pozza2026skew}.  
In particular, starting from the classical Gaussian Laplace approximation, we introduce an explicit skew correction to capture asymmetric posterior features while preserving the computational convenience of Laplace-based methods.
We compare the resulting approximation with the standard Laplace approximation and with MCMC benchmarks. 
Our results show that Laplace-based approximations can be highly effective for posterior density estimation in DPM models, and that the skew-symmetric correction yields a further and systematic improvement, especially when the inferential target is the full posterior distribution rather than posterior point estimates alone. 
These findings suggest that the posterior structure typically encountered in DPM models is not, in itself, incompatible with Laplace approximation.

Our empirical assessment is based on an extensive numerical study combining simulated and real data analyses. 
In the simulation part, we consider four data-generating mechanisms designed to represent different levels of structural complexity: a relatively simple finite Gaussian mixture with four components, a substantially richer mixture with one hundred components and highly heterogeneous weights, and two additional misspecified scenarios based on heavy-tailed kernels. 
For each scenario, we examine a wide range of sample sizes, from $n=20$ up to $n=2{,}000$, in order to assess both finite-sample behavior and scalability. 
Across these settings, we compare the standard Laplace approximation, its skew-symmetric correction, and a slice-sampling benchmark targeting the exact DPM posterior. 
We complement the simulation study with a real-data analysis based on four standard datasets available in \textsf{R}, namely \texttt{faithful}, \texttt{galaxies}, \texttt{iris}, and \texttt{rock}, chosen to illustrate different empirical density shapes and degrees of complexity. 
This combined design allows us to evaluate the proposed methods both under controlled settings, where the truth is known, and on real data, where practical posterior recovery and computational efficiency are the main concerns.

The numerical study evaluates both statistical accuracy and computational efficiency. 
Accuracy is assessed through total variation discrepancies based on posterior density estimates, including comparisons with the true density when available and with a slice-sampling benchmark taken as reference. 
At a finer level, we also investigate how well the proposed approximation reproduces the posterior distribution of the density ordinates across the support. 
Taken together, the results indicate that Gaussian approximations provide a viable route to posterior density estimation in this setting, and that the proposed skew-symmetric correction consistently improves upon the standard Gaussian Laplace approximation while retaining a favorable computational profile relative to fully simulation-based methods.

The remainder of the paper is organized as follows. 
Section~\ref{sec:DPM} is a preliminary section, where we introduce Dirichlet process mixture models, establish the notation used throughout the paper, and briefly review the posterior MCMC methods that are commonly employed in this context. 
Section~\ref{sec:algorithm} is devoted to the proposed approximate sampler. 
Section~\ref{sec:simul} presents the numerical studies. 
Section~\ref{sec:conc} concludes with a discussion and some directions for future work.
Code to reproduce the experiments is available at \if0\blind the GitHub page of the authors.
\else \url{https://github.com/beatricefranzolini/DPLap}.\fi

\section{Dirichlet process mixture models}\label{sec:DPM}
DPM models \citep{lo1984class} are among the most widely used tools in Bayesian nonparametrics for density estimation, clustering, and flexible model-based inference. Their appeal stems from the ability to model complex data-generating mechanisms through mixtures with an infinite number of components. Given observable data $\bm{Y} = (Y_1, \ldots,Y_n)$, a DPM model is defined by 
\begin{equation}
\begin{aligned}
\label{eq:dpm}
Y_i \mid \theta_i &\stackrel{\text{ind}}{\sim} k(\,\cdot\, ; \theta_i), \quad i=1,\dots,n,\\
\theta_i \mid G &\stackrel{\text{iid}}{\sim} G, \quad \quad \quad \;\; i=1,\dots,n,\\
G &\sim \mathrm{DP}(\alpha, G_0),\\
\end{aligned}
\end{equation}
where $k(\cdot;\theta)$ denotes a kernel density with parameter $\theta$, $\mathrm{DP}$ denotes the law of a Dirichlet process \citep{ferguson1973bayesian}, $G_0$ is a baseline probability measure on the parameter space $\Theta$, and $\alpha>0$ is the concentration parameter. 
Integrating out the latent parameters $\theta_1,\dots,\theta_n$, the resulting marginal model for the observations is a random mixture given by
\begin{equation}
f_G(y) = \int_\Theta k(y;\theta)\, G(d\theta).
\label{eq:random-mixture-density}
\end{equation}

A key feature of the Dirichlet process prior is its almost sure discreteness. 
As a consequence, the random measure $G$ admits the almost-sure stick-breaking representation \citep{sethuraman1994constructive}
\begin{equation}
G \overset{a.s.}{=} \sum_{h=1}^\infty \pi_h \, \delta_{\vartheta_h},
\qquad
\vartheta_h \stackrel{\text{iid}}{\sim} G_0,
\label{eq:stick-breaking-G}
\end{equation}
where the law of weights $(\pi_h)_{h\geq1}$ can be described by
\begin{equation}
\pi_h = V_h \prod_{\ell < h} (1 - V_\ell),
\qquad
V_h \stackrel{\text{iid}}{\sim} \mathrm{Beta}(1,\alpha),
\qquad h \geq 1, 
\label{eq:GEM}
\end{equation}
where we use the convention that $\prod_{\ell \in \emptyset} x_{\ell} = 1$ and \eqref{eq:GEM} is denoted for short as $(\pi_h)_{h\geq1}\sim \text{GEM}(\alpha)$.
Substituting \eqref{eq:stick-breaking-G} into \eqref{eq:random-mixture-density}, the induced random density takes the form
\begin{equation*}
f_G(y) \overset{a.s.}{=} \sum_{h=1}^\infty \pi_h \, k(y;\vartheta_h),
\label{eq:dpm-density}
\end{equation*}
which highlights the interpretation of DPM models as countably infinite mixtures. 
The discreteness of $G$ also implies that the latent variables $\theta_1,\dots,\theta_n$ may coincide with positive probability, and, in particular, the marginal probability of ties is $\text{pr}(\theta_i=\theta_j) = (1+\alpha)^{-1}$, for any $i\neq j$ and the conditional probability of tie is $\text{pr}(\theta_i=\theta_j| (\pi_h,\vartheta_h)_{h\geq1}) = \sum_{h\geq1}\pi_h^2$, for any $i\neq j$ \citep[see, for instance,][]{ascolani2024nonparametric,franzolini2025multivariate}. 
This feature of the model induces a random partition of the observations, giving the model its well-known clustering interpretation. 
In light of this, the mixture model in \eqref{eq:dpm} can be conveniently rewritten by introducing a vector of \emph{cluster-label indicators} $\boldsymbol{c}_n = (c_1,\ldots,c_n)$,
\begin{equation*}
\begin{aligned}
Y_i \mid c_i, (\vartheta_h)_{h\geq1} &\overset{\mathrm{ind}}{\sim} 
k(\cdot \mid \vartheta_{c_i})\\
c_i \mid (\pi_h)_{h\geq1} &\overset{\mathrm{iid}}{\sim} \mathrm{Cat}(\pi_1,\pi_2,\ldots),\\
\end{aligned}
\end{equation*}
where $\mathrm{Cat}(\pi_1,\pi_2,\ldots)$ denotes the categorical distribution on the positive integers with probability mass function $\text{pr}(c_i = h \mid (\pi_\ell)_{\ell \geq 1}) = \pi_h$, $\vartheta_h \overset{iid}{\sim} G_0$, for $h \geq 1$, and $(\pi_h)_{h\geq1}\sim \text{GEM}(\alpha)$. 

In the following, we focus on the use of DPM models for density estimation. 
Given observations $Y_1,\dots,Y_n$, the posterior distribution on the random probability measure $G$ induces, through the mapping
\[
G \mapsto f_G, \qquad \text{where}\quad
f_G(y)=\int_{\Theta} k(y;\theta)\,G(d\theta),
\]
a posterior distribution on the unknown sampling density.
In this sense, Dirichlet process mixtures provide a fully nonparametric Bayesian model for density estimation, where inference can be carried out directly on $f_G$ through posterior summaries of the form
\begin{equation*}
\Pi\!\left(f_G \in A \mid Y_1,\dots,Y_n\right),
\qquad A \in \mathcal{B}(\mathcal{F}),
\label{eq:posterior-on-densities}
\end{equation*}
with $\mathcal{F}$ denoting a suitable class of densities. 
A key reason for the widespread success of DPM models in this setting is that the induced prior on $\mathcal{F}$ is highly flexible, allowing the model to adapt to multimodality, skewness, and other departures from simple parametric shapes. 
At the same time, this flexibility is supported by a substantial asymptotic theory: classical results establish posterior consistency for Dirichlet mixtures in density estimation \citep{ghosal1999posterior}, and subsequent refinements derive posterior contraction rates for important classes of mixture priors \citep{ghosal2001entropies,tokdar2006posterior,shen2013adaptive,canale2017posterior}. 
\subsection{Posterior sampling in Dirichlet process mixture models}
Posterior inference via MCMC in nonparametric mixture models is largely popular and is most commonly conducted using one of two broad Gibbs-sampling strategies: \emph{marginal} and \emph{conditional} algorithms. 

Marginal algorithms \citep[see, e.g.,][]{escobar1995bayesian,neal2000markov} integrate out the weights $(\pi_h)_{h\geq1}$ and, in some cases, also the component-specific parameters $(\vartheta_h)_{h\geq1}$, thereby reducing the dimension of the state space and iteratively updating each entry of $\boldsymbol{c}_n$ from its collapsed full conditional. 
In DPM models, these schemes exploit the predictive structure induced by the Dirichlet process \citep{blackwell1973ferguson}. 

Conditional algorithms, by contrast, explicitly retain both weights $(\pi_h)_{h\geq1}$ and component-specific parameters  $(\vartheta_h)_{h\geq1}$ in the Markov chain and update them recursively alongside the allocation indicators in $\boldsymbol{c}_n$ \citep[see, for instance,][]{ishwaran2001gibbs,walker2007sampling,kalli2011slice}.

The choice between these two strategies depends on the specific mixture formulation, the prior specification, and the dimensionality of the problem. 
Marginal algorithms often enjoy improved efficiency from a variance-reduction viewpoint, since collapsing lowers asymptotic variance relative to conditional samplers \citep[see][]{ascolani2025fast}. 
However, while appropriate for quantifying uncertainty of the posterior clustering configuration, the target distribution does not reflect the full uncertainty of the posterior distribution of $G$ and, thus, of $f_{G}$. See \cite{moya2024full} for an explanation and a solution, based on additional sampling steps, to this uncertainty reduction of marginal samplers when the object of inference is density estimation. 
At the same time, conditional samplers admit joint and block updates of latent allocations and component-specific parameters, which may improve global rather than local mixing in multimodal posterior landscapes \citep[see, e.g.,][]{jain2007splitting,ishwaran2001gibbs}. Moreover, they usually involve simpler bookkeeping and are often easier to scale in practice to moderately large datasets \citep[see, for instance,][]{franzolini2026complexity, das2025blocked}.

Among conditional methods, a particularly important distinction is between \emph{blocked} samplers and \emph{slice} samplers. 

Blocked Gibbs samplers rely on truncating the stick-breaking representation of the Dirichlet process as in \citet{ishwaran2001gibbs}. 
This leads to simple conditional updates and a computationally convenient finite-dimensional scheme, at the price of introducing a truncation level.

Slice samplers, initiated in this context by \citet{walker2007sampling} and further developed by \citet{kalli2011slice}, preserve the infinite-dimensional formulation by augmenting the model with auxiliary slice variables instead, so that only a finite random number of components needs to be handled at each iteration, while preserving exact targeting of the posterior. 
Note that, even though the samples obtained are finite at each iteration, since they are from the correct posterior distribution, full posterior uncertainty can be recovered easily, as the distributions of the missing samples are coming from the prior.
Recent work by \citet{franzolini2026complexity} has also clarified their scalability from a theoretical viewpoint, showing that the computational overhead induced by the slice variables grows at most logarithmically with high probability.

Overall, posterior sampling in DPM models is supported by a rich and well-developed MCMC literature, with marginal samplers, blocked Gibbs algorithms, and slice samplers representing the most widely used benchmark strategies. 
These methods provide flexible and accurate inferential tools, but by nature, they still entail a substantial computational burden when the sample size becomes moderately large. 
This computational cost is one of the main motivations for seeking approximate alternatives, as the ones considered in the following.
\section{A new approximate sampler for Dirichlet process mixture models}\label{sec:algorithm}
\subsection{Laplace and skew Laplace approximations}
\label{sec:algorithm:sub1}

In this section, we provide a concise introduction to Laplace \citep[e.g.,][Ch.~13]{gelman2013bayesian} and skew-Laplace \citep{pozza2026skew} approximations of the posterior distribution in Bayesian parametric models. Subsequently, in Section \ref{sec:algorithm:sub2}, we provide the quantities required to apply these techniques to approximate the truncated posterior of Dirichlet process mixture models. 

Let $\bY = (Y_1, \dots, Y_n)$ and
$\theta \in \Theta \subseteq \R^d, d \in \N$ be the parameter of a Bayesian model with posterior distribution 
\begin{equation*}
    p(\theta \mid \bY)
    = 
    \frac{
    p^{un}(\theta \mid \bY)
    }
    {
     \int_{\Theta}
     p^{un}(\theta \mid \bY)
     d \theta
    },
\end{equation*}
where
\begin{equation*}
    p^{un}(\theta \mid \bY) 
    =
    p(\theta)
    p(\bY \mid \theta),
\end{equation*}
denotes the unnormalized posterior, $p(\theta)$ the prior distribution and $p(\bY \mid \theta)$ the likelihood function. Moreover, let
$$
\ell_n (\theta ) 
=
\log
\Big(
p^{un}(\theta \mid \bY)
\Big),
$$ be the log-unnormalized posterior and 
\begin{equation*}
    \ell^{(1)}_n(\theta) =
    \frac{\partial}{\partial \theta} \ell_n(\theta),
    \qquad
    \ell^{(2)}_n(\theta)
    =
    \frac{\partial^2}
    {\partial \theta \partial \theta^{\top}} \ell_n(\theta),
\end{equation*}
be its first and second derivatives.

The Laplace approximation is obtained by taking a second-order Taylor expansion of $\ell_n(\theta)$ around the posterior mode $\tilde{\theta} = \arg \max_{\theta} \ell_n(\theta)$. This yields the following $d$-dimensional Gaussian approximation of $\pi_n(\theta)$:
\begin{equation} \label{eq:lap}
    f_{\textsc{lap}}(\theta) 
    = 
    \frac{1}{(2\pi)^{d/2} \det(\Sigma_{\tilde{\theta}})^{1/2}}
    \exp
    \Big(
    -\frac{1}{2}
    (\theta - \tilde{\theta})^{\top}
    \Sigma_{\tilde{\theta}}^{-1}
    (\theta - \tilde{\theta})
    \Big),
\end{equation}
where $\Sigma_{\tilde{\theta}} = \left[-\ell^{(2)}_n(\tilde{\theta})\right]^{-1}$. Classical results in Bayesian asymptotics demonstrate that, in regular models, $ f_{\textsc{lap}}(\theta) $ converges to $\pi_n(\theta)$ under the total variation distance in probability as $n \to \infty$ \citep{van2000asymptotic}. Moreover, recent theoretical investigations highlight that the Laplace approximation can be accurate even in high-dimensional settings, provided suitable conditions on the interplay between $n$ and $d$ are satisfied \citep{spokoiny2025inexact,katsevich2025improved}. 

While important for justifying the use of the Laplace approximation, these theoretical results are mainly valid from an asymptotic perspective. As a result, in practical settings, it is not uncommon to observe posterior distributions that display important departures from Gaussianity, such as skewness and heavy tails. In recent years, this has motivated substantial interest in developing more flexible posterior approximations, with a particular focus on solutions able to capture skewness \citep[see e.g.,][]{durante2019conjugate,fasano2022scalable,anceschi2023bayesian,katsevich2023laplace,durante2024skewed,zhou2024tractable,dutta2026scalable, kock2026variational}. 

Among these options, we propose to adopt the skew-symmetric correction method developed in \citet{pozza2026skew}. Given any symmetric approximation of the posterior distribution, this technique improves its accuracy by perturbing it with a suitably defined skewness-inducing function. When applied to the Laplace approximation in \eqref{eq:lap}, this method yields the following skew-Laplace approximation of the posterior distribution:
\begin{equation} \label{eq:skew:lap}
    f_{\textsc{s-lap}}(\theta) 
    = 
    2
    f_{\textsc{lap}}(\theta)
    w(\theta),
\end{equation}
with
\begin{equation} \label{eq:skew:fun}
     w_{\tilde{\theta}}(\theta)
    =
    \frac{p^{un}(\theta \mid \bY)}
    {
    p^{un}(\theta \mid \bY)
    +
    p^{un}(2\tilde{\theta} - \theta \mid \bY)
    }.
\end{equation}
Importantly, \eqref{eq:skew:lap} is a proper density function belonging to the skew-symmetric family of distributions \citep{azzalini2003distributions,ma2004flexible}. Moreover, \eqref{eq:skew:fun} does not depend on the posterior normalizing constant. Therefore, \eqref{eq:skew:lap} can be implemented under the mild condition that both the prior and the likelihood are available in closed form. In addition, i.i.d.\ samples from $f_{\textsc{s-lap}}(\theta)$ can be obtained using Algorithm \ref{alg:iid:sampling}, thus allowing direct Monte Carlo evaluation of any functional of $f_{\textsc{s-lap}}(\theta)$. 

\begin{algorithm}
\caption{Skew-Laplace i.i.d sampling} \label{alg:iid:sampling}
\textbf{Input:} $N\in \N$, $\tilde{\theta} \in \R^d, \Sigma_{\tilde{\theta}} \in \R^{d \times d}$ and $w_{\tilde{\theta}}(\theta)$ as defined \eqref{eq:skew:fun}.

\textbf{For} $t = 1,\dots,N$: 

\hspace{5pt}  Sample $\theta'_{t} \sim \mathcal{N}_d(\tilde{\theta},\Sigma_{\tilde{\theta}})$ 

\hspace{5pt}  Set 
$$
\theta_t
=
\begin{cases}
\theta'_t, & \text{with probability }w_{\tilde{\theta}}(\theta'_t), \\
2 \tilde{\theta} - \theta'_t, & \text{otherwise. }
\end{cases}
$$
\textbf{Output:} i.i.d sample $(\theta_1, \theta_2, \dots, \theta_N)$.
\end{algorithm}

From a theoretical point of view, $f_{\textsc{s-lap}}(\theta)$ is guaranteed to provide a better approximation of the posterior than $f_{\textsc{lap}}(\theta)$ under the total variation distance, for any sample size $n$ and any parameter dimension $d$. Moreover, in regular parametric models, $f_{\textsc{s-lap}}(\theta)$ converges to the posterior at a rate that is one order of magnitude faster than $f_{\textsc{lap}}(\theta)$ \citep[see][Sec.~3]{pozza2026skew}.

Since both $f_{\textsc{lap}}(\theta)$ and $f_{\textsc{s-lap}}(\theta)$ are based on a second-order Taylor expansion, they are expected to approximate the posterior distribution well when it is unimodal and sufficiently smooth. These conditions are not necessarily satisfied for truncated posteriors of Dirichlet process mixture models. In the remainder of the paper, we further investigate this issue. First, we derive the quantities required to obtain $f_{\textsc{lap}}(\theta)$ and $f_{\textsc{s-lap}}(\theta)$ for the posterior of Dirichlet process mixture models. Second, we provide empirical evidence that Laplace and skew-Laplace approximations can also be competitive in these settings.

\subsection{Log-posterior and its first and second derivatives in truncated Dirichlet process mixture models}
\label{sec:algorithm:sub2}
To be implemented in practice, the Laplace and skew-Laplace approximations require the posterior mode, the log unnormalized posterior, and its first and second derivatives (the posterior mode is typically obtained using gradient-based optimization methods). In this section, we provide these quantities for the truncated posterior of a Dirichlet process mixture model.

Let $V_h, \theta_h, \pi_h,\, h = 1,2,\dots$ be as defined in Section \ref{sec:DPM}. Moreover, assume that $G_0$ is absolutely continuous with respect to the Lebesgue measure with density $g_0$. Given a truncation level $K \in \N$, let $\alpha, \sigma > 0$, $\bV = (V_1, \dots, V_{K-1})$ and $\btheta = (\theta_1, \dots, \theta_K)$. Then, the unnormalized posterior distribution is equal to 
$$
p(\bV,\btheta \mid \bY)
\propto
\left[\prod_{h=1}^{K-1}(1-V_h)^{\alpha-1}\mathbf 1_{(0,1)}(V_h)\right]
\left[\prod_{h=1}^K g_0(\theta_h)\right]
\prod_{i=1}^n
\left(
\sum_{h=1}^K \pi_h \,\mathcal N(y_i\mid \theta_h,\sigma^2)
\right).
$$

Laplace-based approximations are known to generally perform better on unconstrained parameter spaces. Since $V_h \in (0,1)$, we introduce the following logit reparametrization of $V_h$
$$
R_h=\log\frac{V_h}{1-V_h},
\qquad
V_h=\frac{e^{R_h}}{1+e^{R_h}},
\qquad h=1,\dots,K-1.
$$
For $\bR = (R_1,\dots, R_{K-1})$ the unnormalized posterior becomes
\begin{equation} \label{eq:post:rep}
  p(\bR,\btheta \mid \bY)
\propto
\left[\prod_{h=1}^{K-1}V_h(1-V_h)^{\alpha}\right]
\left[\prod_{h=1}^K g_0(\theta_h)\right]
\prod_{i=1}^n
\left(
\sum_{h=1}^K \pi_h \,\mathcal N(y_i\mid \theta_h,\sigma^2)
\right),
\end{equation}
where the dependence of $V_h, \pi_h, \, h = 1,2, \dots$ is implicit in their definition. Up to an additive constant, the log-posterior in the unconstrained variables $(\bR,\btheta)$ is therefore
$$
\ell(\bR,\btheta)
=
\sum_{h=1}^{K-1}\Bigl[\log V_h+\alpha \log(1-V_h)\Bigr]
+
\sum_{h=1}^K \log g_0(\theta_h)
+
\sum_{i=1}^n \log m_i,
$$
where
$
m_i=\sum_{h=1}^K \pi_h \phi_{ih}
$
with 
$
\phi_{ih}=\mathcal N(y_i\mid \theta_h,\sigma^2).
$
Let $r_{ih}= (\pi_h \phi_{ih})/m_i$ for $i=1,\dots,n $ and $h=1,\dots,K$. Note that $\sum_{h=1}^K r_{ih}=1$ for each $i$. Define for later use
$$
A_{ij} =\frac{\partial \log m_i}{\partial R_j}
=
r_{ij}-V_j\sum_{h=j}^K r_{ih},
$$
for $i=1,\dots,n $ and $j=1,\dots,K.$

The vector of first derivatives of $\ell(\bR, \btheta)$ has elements
\begin{align} \label{eq:gradient}
\frac{\partial \ell(\bR, \btheta) }{\partial R_j}
= &
\,
1-(1+\alpha)V_j
+
\sum_{i=1}^n
\left(
r_{ij}-V_j\sum_{h=j}^K r_{ih}
\right),
\qquad
j \, = \, 1,\dots,K-1,
\nonumber
\\
\frac{\partial \ell(\bR, \btheta)}{\partial \theta_h}
= &
\,
\frac{\partial\ell_{g_0}(\theta_h)}{\partial \theta_h}
+
\sum_{i=1}^n
r_{ih}\left(
\frac{y_i-\theta_h}{\sigma^2}
\right),
\qquad
h \, = \, 1,\dots,K,
\end{align}
where $ \ell_{g_0}(\theta_h) = \log g_0(\theta_h)$.

The hessian matrix of $\ell(\bR, \btheta)$ has instead entries equal to 
\begin{align}
    \frac{\partial^2 \ell(\bR, \btheta)}{(\partial R_j)^2}
     = & \,
     -(1+\alpha)V_j(1-V_j)
     +
     \sum_{i=1}^n
     \Bigl[ (1-2V_j)A_{ij}-A_{ij}^2 \Bigr] 
\nonumber
\\
    \frac{\partial^2 \ell(\bR, \btheta)}{\partial R_j\partial R_k}
    = & \,
    \sum_{i=1}^n
    \left(
    -\,V_jA_{ik}-A_{ij}A_{ik}
    \right),
    \qquad
    \text{for }
    j < k ,
    \nonumber
    \\
    \frac{\partial^2 \ell(\bR, \btheta)}{(\partial \theta_h)^2}
    = & \,
    \frac{\partial^2 \ell_{g_0}(\theta_h)}
    {(\partial \theta_h)^2}
    +
    \sum_{i=1}^n
    \left[
    -\frac{r_{ih}}{\sigma^2}
    +
    r_{ih}(1-r_{ih})
    \left(\frac{y_i-\theta_h}{\sigma^2}\right)^2
    \right],
    \nonumber
    \\
    \frac{\partial^2 \ell(\bR, \btheta)}{\partial \theta_h \partial \mu_k}
    = & \, 
    -\sum_{i=1}^n
    r_{ih}r_{ik}
    \left[
      \frac{(y_i-\theta_h)(y_i-\theta_k)}{\sigma^4}
     \right]
     \qquad
     \nonumber
     \\
     \frac{\partial^2 \ell(\bR, \btheta)}{\partial R_j\,\partial \theta_h}
      = & \,
    \sum_{i=1}^n
    r_{ih}\left(\frac{y_i-\theta_h} {\sigma^2}\right)
    \left(\frac{\partial \log \pi_h}{\partial R_j} -A_{ij} \right),
    \label{eq:hessian}
\end{align}
for $j = 1,\dots, K-1,$ and $ h,k = 1,\dots, K$. The expression of the gradient in \eqref{eq:gradient} can be used within any preferred gradient-based optimization method to identify the maximum $(\tilde{\bR}, \tilde{\btheta})$ of \eqref{eq:post:rep}. Moreover, inverting the negative of the Hessian matrix in \eqref{eq:hessian}, evaluated at $(\tilde{\bR}, \tilde{\btheta})$, yields the covariance matrix of the Laplace approximation. Once the Laplace approximation has been derived, its skew-symmetric corrected version can be obtained following the procedure illustrated in Section \ref{sec:algorithm:sub1}. In Section \ref{sec:simul}, we empirically show that both the Laplace and skew-Laplace approximations provide an accurate characterization of the posterior distribution of Dirichlet process mixture models at a substantially lower computational cost than standard MCMC methods.

\section{Numerical studies}\label{sec:simul}
\subsection{Simulation scenarios and real data}

The numerical study combines controlled simulation experiments with a set of real-data illustrations. 
The simulation component is designed to assess both inferential and approximation accuracy and computational scalability under qualitatively different density structures, whereas the real-data analysis is intended to illustrate the practical behavior of the competing methods on standard benchmark datasets.

\paragraph{Simulation scenario 1.}
The first simulation setting is based on a finite Gaussian mixture with four components, intended to represent a relatively simple but still nontrivial density shape. 
Specifically, data are generated from
\[
Y_i \sim \sum_{h=1}^{4} p_h \, \mathcal{N}(\mu_h, \sigma^2),
\qquad i=1,\dots,n,
\]
with component locations
\[
(\mu_1,\mu_2,\mu_3,\mu_4)=(-3,0,1.5,3),
\]
common standard deviation $\sigma=1$, and weights obtained by a stick-breaking-type construction. 
More precisely, $V_1,\dots,V_4$ are independently sampled from a $\mathrm{Beta}(1,2)$ distribution, and the mixture weights are defined as
\[
p_h = \frac{\prod_{\ell=1}^h V_\ell}{\sum_{j=1}^{4}\prod_{\ell=1}^j V_\ell},
\qquad h=1,\dots,4.
\]
This yields a mixture with moderate multimodality and unequal component masses. 

\paragraph{Simulation scenario 2.}
The second simulation setting is designed to be substantially more challenging. 
Here the data are generated from a Gaussian mixture with $100$ components and common standard deviation $\sigma=1$,
\[
Y_i \sim \sum_{h=1}^{100} p_h \, \mathcal{N}(\mu_h,\sigma^2),
\qquad i=1,\dots,n,
\]
where the component locations are independently sampled as
\[
\mu_h \sim \mathcal{N}(0,1.5^2),
\qquad h=1,\dots,100,
\]
and the weights are set proportional to
\[
w_h = h^{-2},
\qquad h=1,\dots,100,
\]
that is,
\[
p_h = \frac{w_h}{\sum_{j=1}^{100} w_j}.
\]
This corresponds to a Zipf-type weight distribution, yielding a markedly unbalanced mixture with a few dominant components and a long tail of increasingly smaller ones. 
As a consequence, the resulting target density is substantially richer and more irregular than in the first scenario, with many latent components and highly heterogeneous weights. 

\paragraph{Simulation scenario 3.}
The third simulation setting mirrors Scenario~1 in terms of component locations and weights, but replaces the Gaussian kernels with shifted Student-$t$ kernels having $5$ degrees of freedom. 
Specifically,
\[
Y_i \sim \sum_{h=1}^{4} p_h \, t_5(\mu_h),
\qquad i=1,\dots,n,
\]
where $t_5(\mu_h)$ denotes a shifted Student-$t$ distribution with $5$ degrees of freedom centered in $\mu_h$, and the weights $p_1,\dots,p_4$ are sampled as in Scenario~1. 
This setting preserves the same multimodal structure as the first scenario, while introducing heavier tails and therefore a form of model misspecification relative to the Gaussian mixture model used for posterior approximation.

\paragraph{Simulation scenario 4.}
The fourth simulation setting mirrors Scenario~2, again replacing Gaussian mixture components with Student-$t$ kernels with $5$ degrees of freedom. 
Thus,
\[
Y_i \sim \sum_{h=1}^{100} p_h \, t_5(\mu_h),
\qquad i=1,\dots,n,
\]
where the locations $\mu_h$ and the Zipf-type weights $p_h$ are defined exactly as in Scenario~2, 
This produces a highly complex and heavy-tailed target density, combining the large number of latent components and heterogeneous weights of Scenario~2 with additional tail behavior. 
Scenarios~3 and~4 are intended to assess the robustness of the proposed approximations when the data-generating mechanism departs from the Gaussian kernel assumption underlying the fitted DPM model. 

\paragraph{Sample sizes and prior specification}
For all four simulation scenarios, we consider sample sizes
\[
n \in \{20,\,50,\,100,\,200,\,500,\,1000,\,1500,\,2000\},
\]
so as to investigate both finite-sample behavior and scalability. 

In all simulation scenarios, the fitted DPM model uses Gaussian kernels with fixed variance $\sigma^2=1$. 
For all three algorithms and all simulation scenarios, the base measure $G_0$ for the component locations is Gaussian with mean $m_0=0$ and standard deviation $s_0=1$. 
The concentration parameter $\alpha$ of the Dirichlet process is endowed with a $\mathrm{Gamma}(3,3\log(n))$ prior, so that the a priori expected value of $\alpha$ equals $1/\log(n)$.

\paragraph{Real-data illustrations.}
The simulation analysis is complemented by four real-data examples available in \textsf{R}. 
These are the eruption-duration measurements from the \texttt{faithful} dataset, the galaxy-velocity data in \texttt{galaxies}, the \texttt{Petal.Length} variable from the \texttt{iris} dataset, and the \texttt{peri} variable from the \texttt{rock} dataset. 
All four datasets are standardized prior to analysis, and posterior density estimation is carried out on a grid of $400$ equally spaced points covering the observed range of each standardized sample.
 
In these examples, the mixture kernels are Gaussian with fixed standard deviation $0.5$. 
The base measure for the component locations is Gaussian with mean $m_0=0$ and standard deviation $s_0=0.5$, consistently with the standardization of the data and the kernel scale. 
The concentration parameter $\alpha$ is assigned a $\mathrm{Gamma}(3,3)$ prior. 

Since the truth is not available in these examples, the comparison focuses exclusively on agreement with the slice-sampling benchmark, both at the level of posterior mean density estimates and at the level of the posterior distributions of the density ordinates, which is nonetheless the most important metric to evaluate the accuracy of Laplace-based methods.

\subsection{Evaluation metrics}
For each dataset, we run three algorithms: the standard Gaussian Laplace approximation (LAP), the proposed skew-corrected Laplace approximation (Skew-LAP), and the slice sampler (SS) used as simulation-based benchmark. 
The Laplace-based methods are implemented with truncation level $K=20$ for simulated data and truncation level $K=30$ for real data, and generate $2000$ approximate posterior samples, while the slice sampler is run for $10{,}000$ iterations. 
Posterior density estimates are then obtained by evaluating the corresponding sampled mixtures on a common grid, after discarding the first $2000$ slice-sampler draws as burn-in.

The slice sampler, in the improved version presented in Section 3.2 of \cite{ge2015distributed} and available as \texttt{R} code in  \cite{franzolini2026complexity}, is adopted as the reference algorithm for evaluating posterior recovery by the approximate methods considered. 
The rationale for the slice sampler as a benchmark is twofold. 
First, our objective is to assess how closely the proposed approximations reproduce the posterior distribution of the density under the original DPM model. 
This calls for a benchmark that targets the exact posterior of the density itself, rather than the posterior associated with a finite-dimensional truncation, as in blocked Gibbs samplers, or the posterior associated to the clustering configuration, as in marginal samplers. 
Second, the slice sampler scales favorably in practice for DPM models compared to other MCMC methods, and, thus, represents also a competitive alternative in terms of running time \citep{franzolini2026complexity}. 
Consequently, it provides the most natural simulation-based reference against which to assess the quality of the posterior density approximations studied in this work.

We assess the performance of the Laplace and Skew-Laplace algorithms through three complementary discrepancy measures based on total variation distance.

For two densities $f$ and $g$ evaluated on a grid $x_1,\dots,x_N$ with equal spacing $\Delta x$, we approximate their total variation distance by
\begin{equation*}
d^{(N)}_{\mathrm{TV}}(f,g)
=
\frac{1}{2}\sum_{r=1}^N \left| f(x_r)-g(x_r)\right|\,\Delta x .
\label{eq:tv-grid}
\end{equation*}
This quantity measures the global discrepancy between two densities over the whole support: values close to $0$ indicate close agreement, whereas larger values indicate poorer approximation.

Our first criterion compares the posterior mean density estimate produced by method $m$, denoted by $\widehat f_m$, with the true data-generating density $f^\star$, namely
\begin{equation}
d^{(N)}_{\mathrm{TV}}\!\left(\widehat f_m,f^\star\right).
\label{eq:tv-truth}
\end{equation}
This metric evaluates the overall inferential accuracy of the point estimate of the density in simulation settings where the truth is available.

Our second criterion compares the posterior mean density estimate under method $m$ with the corresponding posterior mean density estimate under the slice sampler, that is,
\begin{equation}
d^{(N)}_{\mathrm{TV}}\!\left(\widehat f_m,\widehat f_{\mathrm{SS}}\right).
\label{eq:tv-ss-mean}
\end{equation}
This quantity is, arguably, the most relevant one and should be interpreted as a measure of the goodness of the approximation provided for the density point-estimate relative to the actual posterior. Small values indicate that the approximate method reproduces well the density estimate.

The previous two criteria focus only on point estimates of the density. To assess whether the skew correction also improves the approximation of the posterior uncertainty and, more generally, of the shape of the full posterior distribution of the density, we consider a third metric based on the posterior distribution of the density ordinate at each grid point. 
For a fixed $x_r\in\mathbb{R}$, let $P_{m,r}$ and $P_{\mathrm{SS},r}$ denote the posterior distributions of the scalar $f(x_r)$ under method $m$ and under the slice sampler, respectively. We then define the pointwise posterior discrepancy
\begin{equation}
d_{\mathrm{TV}}\!\left(P_{m,r},P_{\mathrm{SS},r}\right),
\qquad r=1,\dots,N.
\label{eq:tv-pointwise-posterior}
\end{equation}
In practice, the posterior distributions $P_{m,r}$ and $P_{\mathrm{SS},r}$ are available only through Monte Carlo draws, so the distance in \eqref{eq:tv-pointwise-posterior} is computed numerically from their empirical approximations. The collection
\[
\left\{ d_{\mathrm{TV}}\!\left(P_{m,r},P_{\mathrm{SS},r}\right) : r=1,\dots,N \right\}
\]
summarizes how closely method $m$ matches the slice-sampler posterior distribution across the grid. Unlike \eqref{eq:tv-truth} and \eqref{eq:tv-ss-mean}, this criterion does not only assess how well the estimated density is recovered at the point-estimate level, but also captures agreement in posterior spread, local uncertainty quantification, and more generally in the shape of the full posterior distribution of the density ordinate.

\subsection{Results}

\begin{table*}[tbh!]
\caption{Computational time, in seconds, for the three competing methods across the four simulation scenarios and all sample sizes considered. 
Reported times correspond to wall-clock runtime obtained from the \textsf{R} implementations on a laptop with 13th Gen Intel(R) Core(TM) i7-1370P processor. 
For each scenario and sample size, the table reports the runtime of the Gaussian Laplace approximation (Lap), the skew-Laplace approximation (Skew-Lap), and the slice sampler (Slice). 
The table highlights the substantial computational advantage of the Laplace-based approximations relative to the simulation-based benchmark.}
\label{tab:time_sim}
\resizebox{\textwidth}{!}{
\begin{tabular}{l|ccc|ccc|ccc|ccc}
\toprule
&\multicolumn{3}{c}{Simulation scenario 1}&\multicolumn{3}{c}{Simulation scenario 2}&\multicolumn{3}{c}{Simulation scenario 3}&\multicolumn{3}{c}{Simulation scenario 4}\\
\cmidrule(lr){2-4} \cmidrule(lr){5-7} \cmidrule(lr){8-10} \cmidrule(lr){11-13}
Sample size & Lap & Skew-Lap & Slice & Lap & Skew-Lap & Slice& Lap & Skew-Lap & Slice& Lap & Skew-Lap & Slice\\
\midrule
20   & 0.1126  &     0.5829 &  4.1848
&0.1269   &    0.3372 &  3.3820
&0.1176&0.3771&3.9411
&0.1137&0.4212&3.7002
\\
50    & 0.0355   &    0.6130  & 5.2715
&0.0286   &    0.6329 & 5.3831
&0.0352&0.5782&6.3028
&0.0380&0.6252&5.7013
\\
100   & 0.1368   &    0.8994 &  8.2219
&0.0473   &    0.9325 & 7.3032
&0.0487&1.0216&8.9118
&0.0435&0.9588&7.9770
\\
200   &0.2221    &   2.0670  & 11.451
&0.1925   &    2.2633 & 10.652
&0.1979&1.6608&11.616
&0.1735&1.6055&13.944
\\
500   &0.4953    &   4.9542 & 27.518
&0.3692   &    4.5058 & 24.157
&0.4457&4.4416&33.678
&0.2833&3.6463&24.218
\\
1000 &  0.9002    &   9.1795 &  52.186
&0.8433   &    8.7397 & 39.293
&0.7646&8.6205&64.593
&0.6316&7.2151&56.103
\\
1500 & 0.9674   &   12.118 &  79.599
& 0.8434  &    11.518 & 62.610
&1.1452&12.801&84.745
&0.8537&10.558&77.282
\\
2000 & 1.5641  &    16.667 & 100.54
&1.3012   &   16.584 & 81.274
&1.5643&15.581&118.40
&1.5762&14.642&104.67\\
\bottomrule
\end{tabular}}
\end{table*}

\begin{table}[htb!]
\caption{Accuracy summaries for the four simulation scenarios. For each sample size, columns 2--4 report the total variation distance between the posterior mean density estimate produced by Lap, Skew-Lap, and the slice sampler, respectively, and the true data-generating density. 
Columns 5 and 6 report the total variation distance between the posterior mean density estimate of Lap and Skew-Lap, respectively, and the posterior mean density estimate obtained from the slice sampler. 
Smaller values indicate more accurate recovery.
The last column reports the percentage reduction in the TV discrepancy from the slice benchmark achieved by the skew correction relative to the Laplace approximation. }
\label{tab:TV_sim}
\centering
\resizebox{0.75\textwidth}{!}{
\begin{tabular}{lcc|c||ccc}
\toprule
\multicolumn{7}{c}{Simulation scenario 1}\\
\toprule
& \multicolumn{3}{c||}{Truth recovery} & \multicolumn{3}{c}{Point estimate recovery}\\
\cmidrule(lr){2-4} \cmidrule(lr){5-7}
Sample size & Lap & Skew-Lap  & Slice & Lap & Skew-Lap  & \% improvement by Skew \\
\midrule
20  & \textbf{0.0736}    &     0.0762  & \emph{0.0930} & 0.0204   &   \textbf{0.0180} & 12\% \\
50  & \textbf{0.0628}     &    0.0652  & \emph{0.0955} & 0.0593   &   \textbf{0.0453} & 24\%\\
100 & 0.0304     &    \textbf{0.0286}  & \emph{0.0235} & 0.0179   &   \textbf{0.0123} & 31\%\\
200 & \textbf{0.0669}     &    0.0680  & \emph{0.0852} & 0.0391   &   \textbf{0.0306} & 21\%\\
500  & 0.0933    &     \textbf{0.0774} &  \emph{0.0228} & 0.0848  &    \textbf{0.0684} & 19\%\\
1000 & 0.0311    &     \textbf{0.0263} &  \emph{0.0237} & 0.0282  &    \textbf{0.0253} & 10\%\\
1500 & 0.0574    &     \textbf{0.0503} &  \emph{0.0240} & 0.0560  &    \textbf{0.0486} & 13\%\\
2000 & 0.0619    &     \textbf{0.0502} &  \emph{0.0147} & 0.0704  &    \textbf{0.0570} & 19\%\\
\toprule
\multicolumn{7}{c}{Simulation scenario 2}\\
\toprule
& \multicolumn{3}{c||}{Truth recovery} & \multicolumn{3}{c}{Point estimate recovery}\\
\cmidrule(lr){2-4} \cmidrule(lr){5-7}
Sample size & Lap & Skew-Lap  & Slice & Lap & Skew-Lap  & \% improvement by Skew\\
\midrule
20    & \textbf{0.1388}& 0.1541& \emph{0.1983} &0.0663&\textbf{0.0475}&28\%
\\
50   &  \textbf{0.1048}& 0.1082& \emph{0.1163} &0.0234&\textbf{0.0130}&44\%
\\
100   &  \textbf{0.0480}& 0.0582& \emph{0.0557} &0.0247&\textbf{0.0179}&28\%
\\
200   &  \textbf{0.0364}& 0.0382& \emph{0.0336} &0.0097&\textbf{0.0065}&33\%
\\
500  &  \textbf{0.0215}& 0.0239& \emph{0.0234} &0.0036&\textbf{0.0029}&19\%
\\
1000  &  0.0192& \textbf{0.0170}& \emph{0.0191} &0.0076&\textbf{0.0051}&33\%
\\
1500  &  \textbf{0.0138}& \textbf{0.0138}& \emph{0.0133} &0.0049&\textbf{0.0031}&37\%
\\
2000  &  \textbf{0.0201}& 0.0216& \emph{0.0272} &0.0086&\textbf{0.0078}&9\%
\\
\toprule
\multicolumn{7}{c}{Simulation scenario 3}\\
\toprule
& \multicolumn{3}{c||}{Truth recovery} & \multicolumn{3}{c}{Point estimate recovery}\\
\cmidrule(lr){2-4} \cmidrule(lr){5-7}
Sample size & Lap & Skew-Lap  & Slice & Lap & Skew-Lap  & \% improvement by Skew\\
\midrule
20    &\textbf{0.1590}&0.1662&\emph{0.2155}&0.0864&\textbf{0.0737}& 15\%\\
50   &0.0565&\textbf{0.0539}&\emph{0.0488}&0.0334&\textbf{0.0280}& 16\%\\
100   &\textbf{0.0248}&0.0293&\emph{0.0369}&0.0225&\textbf{0.0141}
& 37\%\\
200  &\textbf{0.0654}&0.0715&\emph{0.0942}&0.0458&\textbf{0.0370}
& 19\%\\
500 &0.0826&\textbf{0.0778}&\emph{0.0735}&0.0106&\textbf{0.0086}
 & 19\%\\
1000  &0.0423&\textbf{0.0421}&\emph{0.0423}&0.0149&\textbf{0.0122}& 18\%\\
1500  &0.0506&\textbf{0.0498}&\emph{0.0284}&0.0250&\textbf{0.0231}& 8\%\\
2000  &0.0723&\textbf{0.0673}&\emph{0.0340}&0.0397&\textbf{0.0353}& 11\%\\
\toprule
\multicolumn{7}{c}{Simulation scenario 4}\\
\toprule
& \multicolumn{3}{c||}{Truth recovery} & \multicolumn{3}{c}{Point estimate recovery}\\
\cmidrule(lr){2-4} \cmidrule(lr){5-7}
Sample size & Lap & Skew-Lap  & Slice & Lap & Skew-Lap  & \% improvement by Skew\\
\midrule
20    &0.1894&\textbf{0.1834}&\emph{0.1759}&0.0469&\textbf{0.0313}&33\%\\
50    &0.1126&\textbf{0.0859}&\emph{0.0921}&0.0456&\textbf{0.0303}&34\%\\
100   &0.0383&\textbf{0.0221}&\emph{0.0344}&0.0335&\textbf{0.0208}&38\%\\
200   &0.0962&\textbf{0.0813}&\emph{0.0594}&0.0403&\textbf{0.0269}&33\%\\
500   &0.0676&\textbf{0.0612}&\emph{0.0626}&0.0113&\textbf{0.0079}&30\%\\
1000  &0.0399&\textbf{0.0360}&\emph{0.0434}&0.0153&\textbf{0.0110}&28\%\\
1500  &0.0665&\textbf{0.0600}&\emph{0.0586}&0.0160&\textbf{0.0053}&67\%\\
2000  &0.0589&\textbf{0.0537}&\emph{0.0590}&0.0147&\textbf{0.0104}&29\%\\
\bottomrule
\end{tabular}
}
\end{table}

\FloatBarrier

\begin{figure}[tbh!]
    \centering
    \begin{subfigure}[b]{0.49\linewidth}
        \centering
        \includegraphics[width=\linewidth]{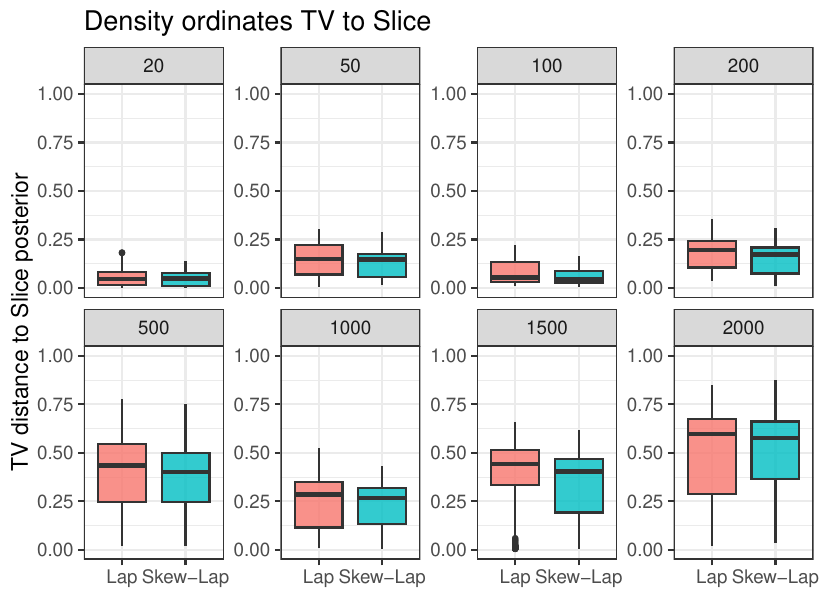}
        \caption{Simulation scenario 1}
        \label{fig:boxplot-sim1}
    \end{subfigure}
    \hfill
    \begin{subfigure}[b]{0.49\linewidth}
        \centering
        \includegraphics[width=\linewidth]{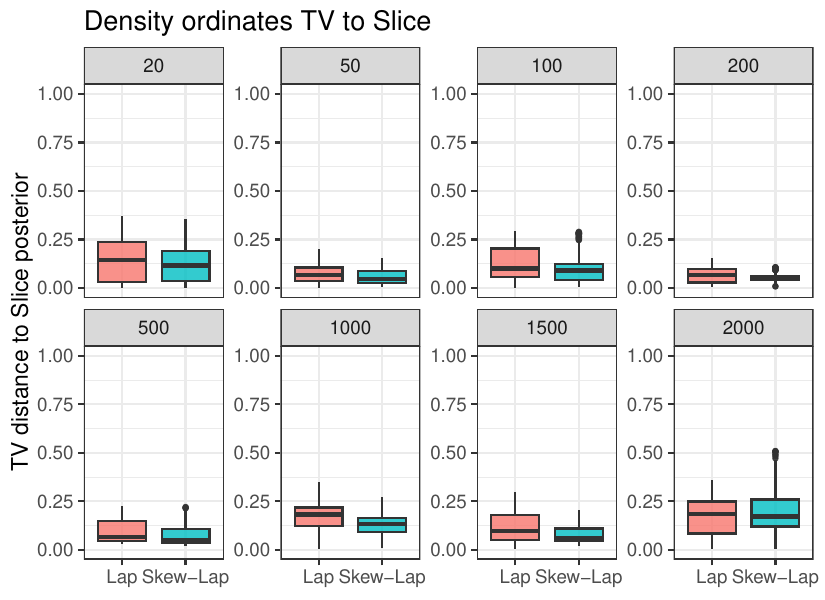}
        \caption{Simulation scenario 2}
        \label{fig:boxplot-sim2}
    \end{subfigure}
    \begin{subfigure}[b]{0.49\linewidth}
        \centering
        \includegraphics[width=\linewidth]{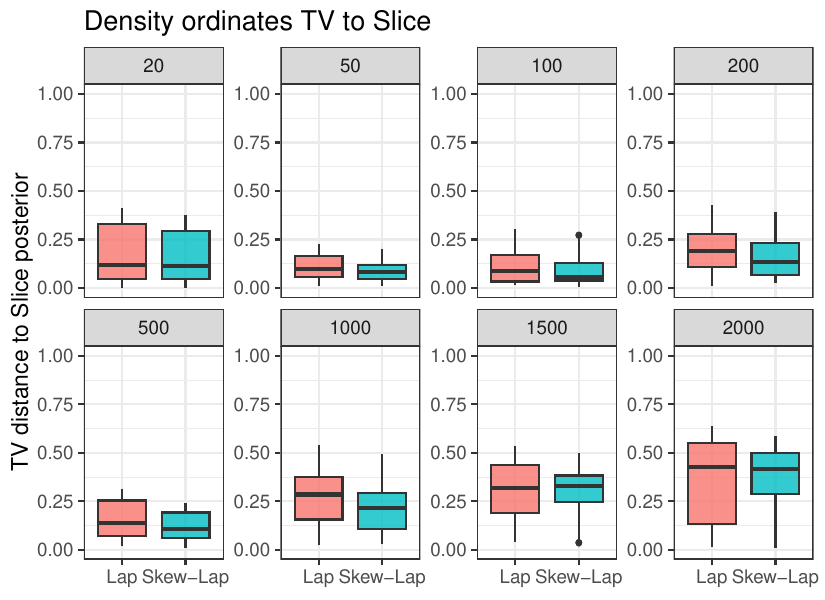}
        \caption{Simulation scenario 3}
        \label{fig:boxplot-sim3}
    \end{subfigure}
    \hfill
    \begin{subfigure}[b]{0.49\linewidth}
        \centering
        \includegraphics[width=\linewidth]{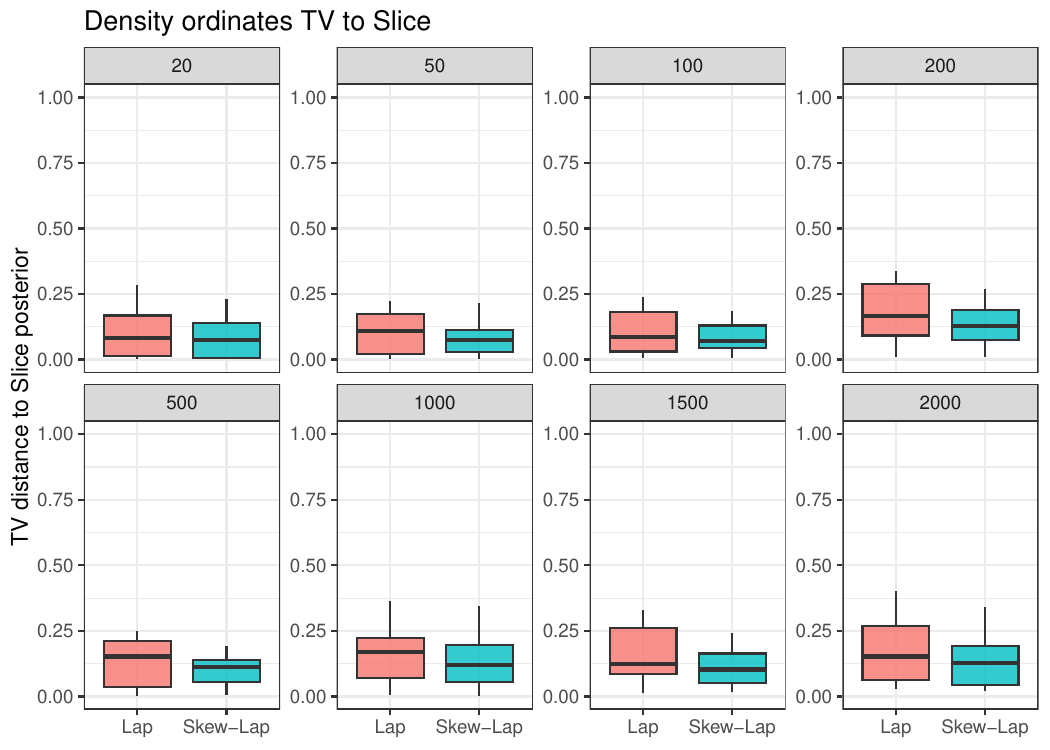}
        \caption{Simulation scenario 4}
        \label{fig:boxplot-sim4}
    \end{subfigure}
    \caption{Pointwise posterior discrepancy with respect to the slice-sampling benchmark in the four simulation scenarios. 
For each sample size and each approximation method (Laplace and Skew-Laplace), the boxplots summarize across grid points the total variation distance between the posterior distribution of the density ordinate $f(x_r)$ under the given approximation and the corresponding posterior distribution under the slice sampler. 
Smaller values indicate closer agreement with the slice-sampler posterior, both in terms of local uncertainty quantification and in the shape of the posterior distribution of the density ordinates. 
Panels (a)--(d) correspond to Simulation Scenarios 1--4, respectively.}
    \label{fig:boxplot_sim}
\end{figure}

The performance assessment and the comparison between the two Laplace-based methods is developed along three main dimensions: computational efficiency, accuracy of posterior mean density recovery, and fidelity in approximating the posterior distribution of the density curve.

Table~\ref{tab:time_sim} shows that both Laplace-based procedures are consistently and substantially faster than the slice sampler across all scenarios and sample sizes. 
The standard Laplace approximation is the fastest method overall. 
The skew correction entails a moderate additional computational cost, roughly a tenfold increase relative to the plain Laplace approximation, but it remains far less expensive than the slice sampler in every setting. 
Importantly, this gap widens as the sample size increases: at $n=2000$, the slice sampler requires between $80$ and $120$ seconds, depending on the scenario, compared with about $1.5$ seconds for Laplace and $15.5$ seconds for Skew-Laplace. 
These results are in line with expectations and suggest that Laplace and Skew-Laplace methods may be computationally feasible, in terms of running time, even in higher dimensional settings where MCMC-based inference is impractical.

Table~\ref{tab:TV_sim} presents the results in terms of posterior mean density recovery in the four simulation scenarios.  
When accuracy is measured against the true data-generating density, the two approximations perform similarly, and neither dominates uniformly. 
This is an unsurprising finding, since a good posterior approximation need not translate directly into better point estimation. 
A clearer pattern emerges when the slice-sampler posterior mean is used as the benchmark. 
In this case, the skew correction consistently improves upon the Gaussian Laplace approximation across all configurations, with percentage reductions in total variation distance typically falling in the 15–40\% range. 
The improvement is most pronounced in Scenarios~2 and 4, where the percentage reduction is at least 28\% in 14 of the 16 datasets. 
By contrast, in Scenarios~1 and 3, 14 of the 16 datasets show an improvement of no more than 24\%.
Since Scenarios~2 and 4 involve 100 mixture components and Zipf-type weights, whereas Scenarios~1 and 3 are generated from a true density with only four components, these results suggest that the skew correction is particularly effective in more complex density settings.

Figure~\ref{fig:boxplot_sim} shows full posterior recovery for the density ordinates and therefore provides the most direct assessment of whether the approximations reproduce not only posterior point estimates but also the full posterior distribution of the density. 
The boxplots show that Skew-Laplace consistently yields lower or comparable pointwise total variation distances to the slice-sampler posterior across all four scenarios. 
The advantage is again particularly visible in Scenarios~2 and~4, where the density structure is most complex and heavy-tailed. 
For the interpretation of Figure~\ref{fig:boxplot_sim} is worth emphasizing a small technical note: the main information conveyed by these boxplots lies in the relative comparison between Laplace and Skew-Laplace, rather than in the absolute magnitude of the pointwise TV values themselves. 
Since these discrepancies are computed for the posterior distribution of the scalar ordinate $f(x_r)$ at each grid point, their numerical value may depend on the local scale of the density: in regions where the density is very small, even a practically modest shift in the posterior distribution of $f(x_r)$ may induce a visible TV discrepancy from the Slice posterior. 
Accordingly, the most meaningful feature of Figure~\ref{fig:boxplot_sim} is the systematic downward shift of the Skew-Laplace boxplots relative to their Gaussian counterparts, indicating a closer match to the slice-sampler posterior. 
This is an important finding: the primary motivation for the skew correction is precisely to recover posterior asymmetry, which do not result only in a possible shift the posterior mean, but should also for a more robust uncertainty recovery, and Figure~\ref{fig:boxplot_sim} confirms that this objective is at least partially achieved in all considered scenarios.

\begin{table}[tbh!]
\caption{Results for the real-data examples. 
For each dataset, the table reports the sample size, the computational time in seconds for Lap, Skew-Lap, and the slice sampler, and the total variation distance between the posterior mean density estimate of each approximation method and the posterior mean density estimate obtained from the slice sampler. 
Since the true density is unavailable for real data, agreement is evaluated only relative to the slice-sampling benchmark. 
The last column reports the percentage reduction in the TV discrepancy from the slice benchmark achieved by the skew correction relative to the Gaussian Laplace approximation.}
\label{tab:result_real}
\resizebox{\textwidth}{!}{
\begin{tabular}{lc|ccc||ccc}
\toprule
&& \multicolumn{3}{c||}{Time in sec} & \multicolumn{3}{c}{Point estimate recovery} \\
\cmidrule(lr){3-5} \cmidrule(lr){6-8}
Data & Sample size & Lap & Skew-Lap & Slice & Lap & Skew-Lap & \% improvement by Skew\\ 
\midrule
\texttt{faithful}  & 272  & 0.6605    &   4.0232  & 21.594  
&  0.0709  &   \textbf{0.0628} & 11\%\\
\texttt{galaxy} & 82  &  0.1572  &  1.3964 & 9.1154  
&  0.0541  &   \textbf{0.0420} & 22\%\\
\texttt{iris} & 150  &  0.2276   & 2.540 & 13.964  
&  0.0441  &   \textbf{0.0421} & 5\%\\
\texttt{rock} & 48  & 0.0862   &    0.8589 & 8.1719 
& \textbf{0.0358}  &   \textbf{0.0358} & 0\% \\
\bottomrule
\end{tabular}
}
\end{table}

\begin{figure}[tbh!]
    \centering
     \includegraphics[width=\linewidth]{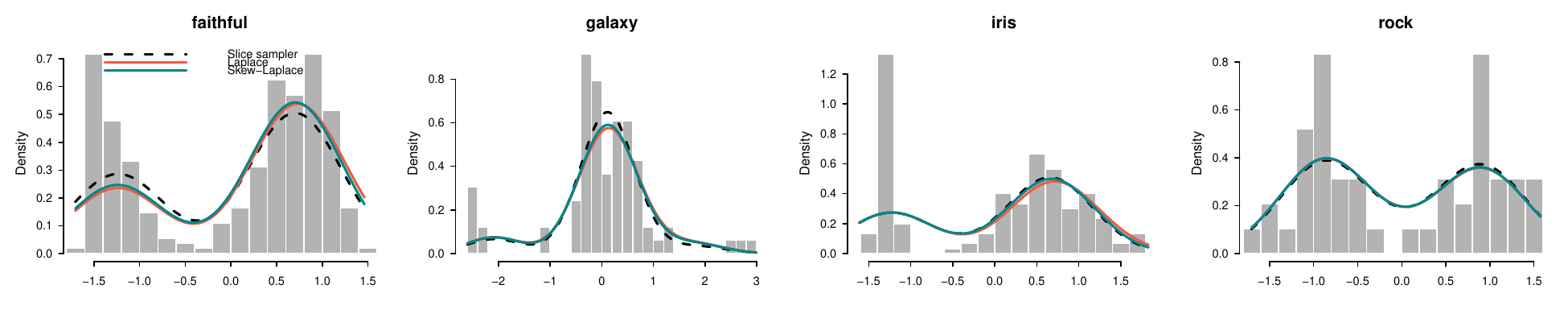}
    \caption{Posterior mean density estimates for the real-data examples. 
Histograms represent the standardized observed data, while the superimposed curves correspond to the posterior mean density estimates obtained from the slice sampler (black dashed line), the Gaussian Laplace approximation (red solid line), and the Skew-Laplace approximation (light-blue solid line). 
The four panels correspond to the \texttt{faithful}, \texttt{galaxies}, \texttt{iris}, and \texttt{rock} datasets, respectively. 
The figure compares the point estimates from the competing methods with the simulation-based benchmark.}
    \label{fig:density-real}
\end{figure}

\begin{figure}[tbh!]
    \centering
     \includegraphics[width=0.4\linewidth]{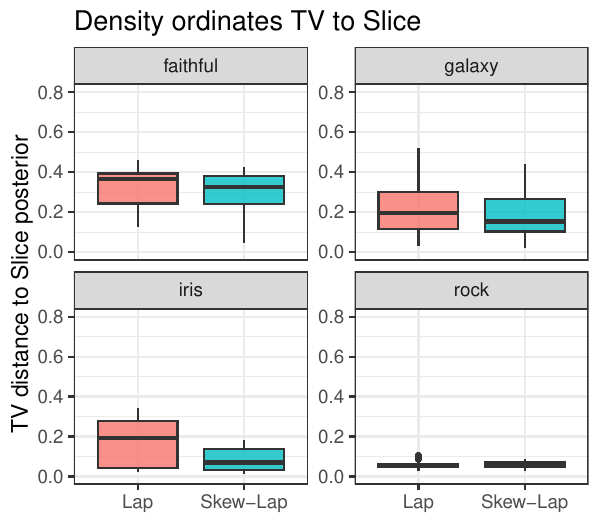}
     \caption{
For each dataset and for each approximation method (Laplace and Skew-Laplace), the boxplots summarize across grid points the total variation distance between the posterior distribution of the density ordinate under the approximation and the corresponding posterior distribution under the slice sampler. }
    \label{fig:boxplot-real}
\end{figure}

Table~\ref{tab:result_real} and Figures~\ref{fig:density-real} and \ref{fig:boxplot-real} refer to results on real data and broadly confirm the simulation findings. 
Both Laplace-based methods remain far faster than the slice sampler on all four datasets. 
The skew correction improves agreement with the slice-sampler posterior mean in three out of four cases, with the largest gains on the \texttt{galaxy} (22\%) and \texttt{faithful} (11\%) datasets. The gain is modest for \texttt{iris} and neutral for \texttt{rock}. 
Figure~\ref{fig:boxplot-real} further shows that the skew-corrected approximation generally yields a more accurate recovery of the full posterior distribution of density ordinates, reinforcing the same pattern seen in the simulation study.

Taken together, these results suggest that Gaussian Laplace approximation is a viable and underexplored approach to posterior density estimation in DPM models, and that the skew-symmetric correction provides a meaningful and consistent refinement, particularly for complex density structures.

\section{Discussion and future directions}\label{sec:conc}

This work has investigated Laplace-based posterior approximation for density estimation in Dirichlet process mixture models, with particular emphasis on a skew-symmetric correction designed to enrich the classical Gaussian approximation. 
The results suggest, first, that Gaussian Laplace approximations are more viable in this setting than might have been expected. 
The posterior structure of DPM models, shaped by clustering uncertainty, label switching, and effective multimodality, might appear fundamentally unfavorable to local Gaussian approximations, and this may partly explain why such methods have received limited attention in the Bayesian nonparametric literature. 
The numerical evidence reported here suggests that this skepticism is at least partly excessive. 
Across a broad range of simulated and real-data settings, the Gaussian Laplace approximation provides a reasonably accurate first-order description of the posterior distribution at a small fraction of the computational cost required by MCMC. 

In addition, the proposed skew-symmetric correction provides a meaningful and consistent across-scenarios refinement of this baseline approximation. 
In particular, the skew-corrected approximation improves systematically on the Gaussian Laplace approximation in recovering the slice-sampler posterior mean density and in reproducing the full posterior distribution of the density ordinates, especially in scenarios with richer underlying density structure.

From a computational perspective, the current implementation shows that the skew correction introduces a non-negligible additional cost relative to the plain Gaussian Laplace approximation, although it remains much faster than the slice sampler in all configurations considered. 
The exact gap between the two approximate methods, however, partly reflects implementation choices rather than an intrinsic limitation of the methodology. 
In the present code, the Gaussian Laplace approximation relies heavily on highly optimized low-level routines already available in \textsf{R}, such as \texttt{optim} for mode finding and \texttt{mvtnorm::rmvnorm} for Gaussian sampling, whereas the additional skew-sampling step is currently implemented directly in interpreted \textsf{R} code. 
The computational profile reported here is therefore likely conservative, and a more optimized implementation of the skew correction, for example, in a lower-level language, could substantially reduce the current overhead and make the skew-corrected approximation even more competitive in practice.

Some limitations of the present study should also be acknowledged. 
All experiments are conducted in univariate settings, i.e., with scalar data, which, while they are probably the most used setting for controlled algorithmic experiments for Bayesian mixtures \citep[see e.g.,][]{fruhwirth2004estimating,papaspiliopoulos2008retrospective,das2025blocked,ascolani2025fast,franzolini2026complexity}, taken alone, are a substantial simplification relative to many applications of DPM models. 
Whether the same conclusions on Laplace-based approaches continue to hold in multivariate problems remains a direction to investigate further. 
Moreover, our contribution is based on numerical evidence; while the simulation studies are extensive, the theoretical properties of the skew-symmetric approximation in the DPM setting remain largely unexplored. 
In particular, it is not yet clear how its accuracy depends on posterior concentration, asymmetry, and possible model misspecification. These issues deserve a dedicated theoretical treatment.

One especially promising direction for future work concerns clustering recovery. 
The focus of the present paper is on posterior density estimation, but the approximate posterior samples produced by both Laplace and skew-Laplace methods can also be used to generate Monte Carlo samples of the latent clustering configuration, given the draws of the truncated $G$. 
This is potentially important, since even moderate differences in posterior uncertainty for the component-specific parameters and weights may propagate into more visible differences in clustering behavior and summaries.
From this perspective, the advantages of the skew correction may prove even more pronounced for clustering than for density estimation alone, although this remains to be assessed in a separate study.

Another direction seems natural, i.e.\ to investigate whether the same approximation strategy can be extended beyond Dirichlet process mixtures to related Bayesian nonparametric constructions, such as hierarchical Dirichlet processes \citep{teh2006hierarchical} or Pitman--Yor mixtures \citep{pitman1997two}, where exact posterior simulation is often even more demanding and where accurate approximate alternatives may therefore be especially valuable.

Overall, the results of this paper suggest that Laplace-based posterior approximation may deserve greater attention in Bayesian nonparametric mixture modeling than it has received so far. 
The Gaussian approximation already provides a useful and surprisingly effective baseline for density estimation, while the proposed skew-symmetric correction offers a meaningful and often practically relevant improvement. 
Taken together, these findings point to a promising middle ground between crude deterministic approximation and expensive simulation-based inference.

\subsection*{Acknowledgments}
Francesco Pozza is funded by the European Union (ERC, PrSc-HDBayLe, n. 101076564). 


\bibliographystyle{cas-model2-names}

\bibliography{biblio}



\end{document}